\documentclass[aps,twocolumn,preprintnumbers,floatfix,nofootinbib]{revtex4-1}
\pdfoutput=1
\usepackage{graphicx}
\usepackage{bm}
\usepackage{times}
\usepackage{slashed}
\usepackage{color}
\usepackage{slashed}
\usepackage{lipsum}
\usepackage{subfigure}
\usepackage[utf8]{inputenc}
\usepackage{amsfonts}
\usepackage{mathrsfs}
\usepackage{amsmath}
\usepackage{graphicx,epsfig}
\usepackage{url}
\usepackage{amssymb}
\usepackage{comment}
\newcommand{\be}{\begin{equation}}
\newcommand{\ee}{\end{equation}}
\newcommand{\bea}{\begin{eqnarray}}
\newcommand{\eea}{\end{eqnarray}}
\usepackage{hyperref}

\begin{document}

\title{Is the dark matter particle its own antiparticle?}

\author{Farinaldo S. Queiroz}
\email{farinaldo.queiroz@mpi-hd.mpg.de}

\author{Werner Rodejohann}
\email{werner.rodejohann@mpi-hd.mpg.de}

\author{Carlos E. Yaguna}
\email{carlos.yaguna@mpi-hd.mpg.de}

\affiliation{Max-Planck-Institut f\"ur Kernphysik, Postfach 103980, 69029 Heidelberg, Germany}

\begin{abstract}
We propose a test based on direct detection data that allows to determine if the dark matter particle is different from its antiparticle. The test requires the precise measurement of the dark matter spin-independent direct detection cross sections off \emph{three} different nuclei, and consists of interpreting such signals in terms of self-conjugate ($\mbox{particle} = \mbox{antiparticle}$) dark matter to see if such interpretation is consistent. If it is not, the dark matter must be different from its antiparticle.  We illustrate this procedure for two sets of target nuclei, $\mathrm{\{Xe, Ar, Si\}}$ and $\mathrm{\{Xe, Ar, Ge\}}$,  identifying the regions of the parameter space where it is particularly feasible.  Our results indicate that future signals in direct detection experiments, if sufficiently accurate, might be used to establish that the dark matter particle  is not  its own  antiparticle --a major step towards the determination of the fundamental nature of the dark matter.

\end{abstract}


\maketitle

{\bf Introduction.} 
The existence of dark matter (DM) has been established, via its gravitational effects,  through a variety of observations at different scales --from galaxies to the largest structures in the Universe. According to the most recent data, DM accounts for about $27\%$ of the total energy density of the Universe and about $85\%$ of its matter density \cite{Ade:2015xua}. And yet,  the fundamental nature of the DM particle (e.g. its mass, spin, quantum numbers, etc.) remains a mystery, providing one of the most important open problems in particle and  astroparticle physics today \cite{Bertone:2004pz, Bertone:2010zza}. Its solution will be  crucial, in particular, to establish the new Standard Model of particle physics.

Within this context, it is essential to determine whether the DM particle  is its own antiparticle. This amounts to whether it is a Majorana or a Dirac particle for fermion DM,  or whether  it is a real or a complex particle for scalar and vector DM.  Among the multitude of models that have been proposed to account for the DM, it is not difficult to find suitable candidates for  each of these cases --see e.g. \cite{Jungman:1995df,Primack:1988zm, Silveira:1985rk, Arina:2007tm, Servant:2002aq, Hambye:2008bq}. The neutralino of the MSSM \cite{Jungman:1995df} and a heavy neutrino \cite{Belanger:2007dx} supply,  for example,  well-known instances of Majorana and Dirac DM, respectively. But, is there a way to \emph{experimentally} distinguish between these two possibilities?

\begin{figure}[tb!]
\begin{center}
\includegraphics[width=0.5\columnwidth]{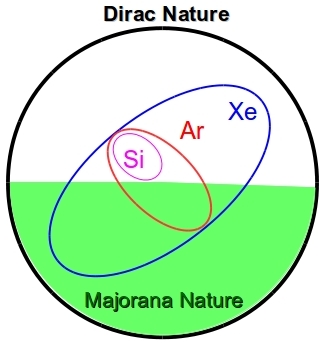}
\caption{\small An illustration of the test we are proposing: after combining direct detection data from three different nuclei (e.g. Xe, Ar, and Si)  the Dirac nature of the DM could be singled out. } 
\label{DMnature}
\end{center}
\end{figure}

Here, we propose a test that allows to determine if the DM particle is different from its antiparticle. It relies on direct detection data only and requires  the precise measurement of the DM spin-independent cross section off three different nuclei. The idea is to interpret these measurements in terms of self-conjugate DM ($\mbox{particle}=\mbox{antiparticle}$) and see if that interpretation is consistent. If it is not, the DM particle must be different from its antiparticle.   In other words, it is a test that can exclude  a DM particle that is identical to its antiparticle. Figure \ref{DMnature} illustrates pictorially how the test works.

For clarity, in our discussion we will initially focus on fermion DM and the possibility of differentiating  between  Dirac and  Majorana DM, but we later show that the test  works the same way for scalar and vector DM.  

{\bf Spin-independent cross sections.} For a fermion DM particle, $\chi$, the most general Lagrangian  describing its spin-independent (SI) interactions with point-like nucleons ($N=p,n$) is given by \cite{Belanger:2008sj} 
\begin{equation}
\mathscr{L}_{_{SI}}^F = \lambda_{N,e}\, \bar\psi_\chi\psi_\chi\,\bar\psi_N\psi_N
+\lambda_{N,o}\, \bar\psi_\chi\gamma_\mu\psi_\chi\,\bar\psi_N\gamma^\mu\psi_N\,,
\end{equation}
where $\lambda_{N,e}$ and $\lambda_{N,o}$ are dimensionful couplings that are determined by the underlying particle physics model and, for simplicity, are taken  here to be real. The indices $e$ and $o$ indicate whether the corresponding operator is even or odd with respect to $\chi\leftrightarrow\bar\chi$ interchange. For a Majorana DM particle only the even operators are non-zero and the DM elastic scattering cross section off a point-like nucleus takes the simple form
\begin{equation}
\label{eq:sigmaM}
\sigma^M_{SI}=\frac{4\mu_\chi^2}{\pi}\left[\lambda_{p}^M\,Z+\lambda_{n}^M\,(A-Z)\right]^2\,,
\end{equation}
where $\lambda_{N}^M\equiv \lambda_{N,e}$, $ \mu_\chi=M_\chi M_A/(M_\chi+M_A)$ is the WIMP-nucleus reduced mass, $Z$ is the nucleus charge, and $A$ is the total number of nucleons. For a Dirac DM particle, the even and odd operators are in general both present and there are two different spin-independent cross sections with nuclei --one for the DM particle  ($\sigma_{SI}$),  and one for the DM antiparticle ($\bar \sigma_{SI}$).  It is useful to define, in this case, the parameters $\lambda_N^D\equiv(\lambda_{N,e}+\lambda_{N,o})/2$ and $\lambda_N^{\bar D}\equiv(\lambda_{N,e}-\lambda_{N,o})/2$, which respectively determine $\sigma_{SI}$ and $\bar \sigma_{SI}$. Within the standard WIMP framework,   the DM particles and antiparticles are expected to have the same  or very similar densities, so the experimentally relevant direct detection cross section for a Dirac DM particle, $\sigma_{SI}^D$,  is simply the average between $\sigma_{SI}$ and $\bar\sigma_{SI}$,
\begin{align}
\label{eq:sigmaD}
\sigma_{SI}^D =\frac{4\mu_\chi^2}{\pi}\frac 12 &\left( \left[\lambda_p^D\,Z+\lambda_n^D\,(A-Z)\right]^2 \right. \\
 & + \left. \left[\lambda_p^{\bar D}\,Z+\lambda_n^{\bar D}\,(A-Z)\right]^2 \right).\nonumber
\end{align}

For definiteness,  the spin-dependent contribution to the signal rate is taken to be negligible throughout our analysis. The event rate at a DM detector is  then given by $R=\sigma_X I_X$, where $\sigma_X$ is the Dirac or Majorana cross section off the nucleus $X$ --equations (\ref{eq:sigmaM}) or (\ref{eq:sigmaD})--, which is determined by the underlying particle physics model;  $I_X$, on the other hand, depends on experimental, astrophysical, and nuclear physics inputs. Explicitly, 
\begin{equation}
I_X= n_\chi N_T\int dE_R\int_{v_{min}}^{v_{max}}d^3v\, f(v) \frac{m_X}{2v\mu_X^2}F_X^2(E_R)\,,
\end{equation}
where  $n_\chi$ is the local  number density of DM particles, $N_T$ is the number of target nuclei, $v$ is the velocity of the DM particle, $f(v)$ is the DM velocity distribution, and $F_X$ is the nuclear form factor. Usually, $I_X$ is assumed to be given 
(e.g. by assuming the standard halo model --see however \cite{Duda:2006uk, Kelso:2016qqj, Sloane:2016kyi, Bozorgnia:2016ogo} for alternative approaches), allowing to translate the absence of  signals in current experiments  into upper bounds on the  cross sections \cite{Aprile:2012nq, Akerib:2016vxi}. In the same way, an observed signal in a future experiment could be translated into a certain range for the cross section.  

{\bf Dirac or Majorana dark matter?} Since the direct detection cross sections for Dirac and Majorana DM have different functional dependencies, it should be possible to determine, given a certain number of observations, whether the experimental data is described by equation (\ref{eq:sigmaM}) or equation (\ref{eq:sigmaD}). That is, it seems feasible to distinguish  Dirac DM from Majorana DM with just direct detection data. Our proposal to do so is to try to explain the direct detection signals in terms of Majorana DM, and check if that interpretation is consistent. If it is not, it can be ascertained that the DM is a Dirac particle.

To illustrate the basic idea behind this test, we first consider a simplified scenario. Let us assume, for example, that  the DM cross section off the nuclei ${^{A_X}_{Z_X}X}$ and ${^{A_Y}_{Z_Y}Y}$ have been measured to be  $\tilde \sigma_{X}$ and $\tilde\sigma_Y$, respectively. Interpreting these results in terms of Majorana DM implies, from equation (\ref{eq:sigmaM}), that
\begin{align}
\label{eq:sigmaX}
\left[\lambda_p^M\,Z_{X}+\lambda_n^M\,(A_{X}-Z_{X})\right]^2=& \frac{\pi \tilde \sigma_{X}}{4 \mu_\chi^2},\\
\left[\lambda_p^M\,Z_{Y}+\lambda_n^M\,(A_{Y}-Z_{Y})\right]^2=& \frac{\pi \tilde \sigma_{Y}}{4 \mu_\chi^2}.
\end{align}
Each of these equations describes, in the plane ($\lambda_{p}^{M},\lambda_{n}^{M}$), two parallel lines  with slopes $m_X=Z_X/(A_X-Z_X)$ and $m_Y=Z_Y/(A_Y-Z_Y)$.  For $m_X\neq m_Y$ these  lines would always intersect, in pairs, at four different points. These four values of ($\lambda_p^M$, $\lambda_n^M$) would thus be consistent with both measurements (notice though that  two of these solutions correspond to the global sign change $(\lambda_p^M, \lambda_n^M)\to (-\lambda_p^M, -\lambda_n^M)$, so only two solutions are actually different). Thus, two DM scattering signals off different nuclei are always consistent with Majorana DM.  But, if the DM cross section off another nuclei, say $V$, were also measured  and found to  be $\tilde \sigma_V$, two things could happen with its corresponding parallel lines:
\begin{enumerate}
\item At least one of them would pass through one of the two distinct points consistent with the $X$ and $Y$ measurements. In this case, we would have determined, up to a global sign,  the values of $\lambda_p^M$ and $\lambda_n^M$, but we could not say anything about the Dirac or Majorana nature of the DM particle.

\item None of them would pass through one of the two distinct points consistent with the $X$ and $Y$ measurements. In this case, the three measurements would be inconsistent and our interpretation in terms of a Majorana DM particle must be wrong. We would conclude, therefore, that the DM is  a Dirac particle. 
\end{enumerate}

Thus, direct detection signals can be used, at least in principle, to exclude  the possibility of Majorana DM, establishing the Dirac nature of the DM particle. But such signals could never demonstrate the Majorana nature of the DM particle or, equivalently, exclude a Dirac DM particle. The reason for this asymmetry is that the  Dirac DM detection cross section can always be reduced to a Majorana form, but not the other way around. At the Lagrangian level the issue is clear: a Dirac DM particle has both scalar and vector interactions whereas a Majorana DM particle  has only scalar interactions.

To be able to say anything about the Dirac or Majorana nature of the DM particle, the observed signals must therefore be due to a Dirac particle. In that case, our problem consists in determining whether, for a given set of $\{\lambda_{p}^D,\lambda_{p}^{\bar D},\lambda_{n}^D,\lambda_{n}^{\bar D}\}$, there exist values of $\lambda_{p}^M$ and $\lambda_{n}^M$ such that the system 
\begin{equation}
\label{eq:system}
\sigma_{SI,k}^M=\sigma_{SI,k}^D\quad \mathrm{for~} k=1,2,3
\end{equation}
(where $k$ denotes the different nuclei) is inconsistent. Notice that the reduced mass drops out from this equation and with it  the dependence on the DM mass. Thus, whether the  system in equation (\ref{eq:system}) is inconsistent or not does not depend on the mass of the DM particle. By the same token, one can see that this system is invariant under a global rescaling of all the couplings, making the precise value of the cross sections irrelevant to the issue of Dirac vs.\ Majorana DM.  Finally, this system depends on the number or protons and neutrons in the nuclei only through their ratio, $Z/(A-Z)$, which, as we have already argued, should take  different values for the three nuclei considered. It turns out though  that $Z/(A-Z)$ varies only between $1$ and $0.65$ for stable nuclei, and it is  very similar for some of the nuclei that are of interest to DM experiments: about $1$ for $\mathrm{\{Si,Ca,O\}}$, about $0.68$ for $\mathrm{\{Xe,W\}}$ and about $0.91$ for $\mathrm{\{F,Na\}}$. To help  discriminate between Dirac and Majorana DM, only one element of each of these sets is actually useful.

Before proceeding, let us emphasize that there are some \emph{special} cases in which direct detection experiments cannot  differentiate between Dirac and Majorana DM, because the Majorana direct detection cross section takes exactly the same form as the Dirac one, namely: (i) When the Dirac fermion has  either scalar or vector interactions but not both; (ii) When the couplings of the Dirac fermion are such that either the particle or the antiparticle cross section vanishes for all nuclei or the density of the Dirac dark matter particle is much smaller or larger than the density of the antiparticle; (iii) When the DM particle couples only to protons or neutrons  but not to both; (iv) When the ratio between the coupling to the proton and to the neutron is the same for the DM particle and the antiparticle.  

\renewcommand{\arraystretch}{1.2}
\begin{table}[tb!]
\begin{center}
\begin{tabular}{c c c}
	\hline
	$\mathrm{_{54}Xe}$	& $\mathrm{_{32}Ge}$ & $\mathrm{_{14}Si}$\\
    \hline 
   128 (1.9\%) & 70 (21\%) & 28 (92\%)\\
   129 (26\%) & 72 (28\%) & 29 (4.7\%)\\
   130 (4.1\%) & 73 (4.7\%) & 30 (3.1\%)\\
   131 (21\%) & 74 (36\%) & ~\\
   132 (27\%) & 76 (7.4\%) & ~\\
   134 (10\%) & & \\
   136 (8.9\%) & &
\end{tabular}
\caption{\small The value of $A$ for the different isotopes of $\mathrm{Xe}$, $\mathrm{Ge}$ and $\mathrm{Si}$, with their respective fractional abundances \cite{Feng:2011vu}. There are no relevant isotopes for Ar.\label{tab:isotopes}}
\end{center}
\end{table}

{\bf Application.} Let us now go beyond the simplified framework introduced above and consider a more realistic setup. First of all, the scattering cross sections can only be measured up to a certain accuracy ($\Delta\sigma$). Thus, $\tilde \sigma_X$ in  equation (\ref{eq:sigmaX}) should be replaced by $\tilde \sigma_X\pm \Delta \tilde\sigma_X$, with the result that the straight-line solutions actually consists of bands of finite width, and the intersecting points become overlapping regions. Clearly, the larger the uncertainty in the cross sections the more difficult it will be  to distinguish Dirac from Majorana DM. Regarding the expected uncertainty on the determination of the cross sections, several  previous works have already  estimated its size under different conditions \cite{Pato:2010zk,Roszkowski:2016bhs}. An accuracy of order $10\%$ seems to be achievable for the fiducial astrophysical setup \cite{Pato:2010zk}  and a light DM particle, $M_\chi\lesssim 100$ GeV. To simplify the analysis, throughout the rest of the paper the   DM mass is set to $100$ GeV,  and  the accuracy  in the determination of the cross sections is taken to be the same for all targets.

\begin{figure}[tb!]
\begin{center}
\includegraphics[width=0.9\columnwidth]{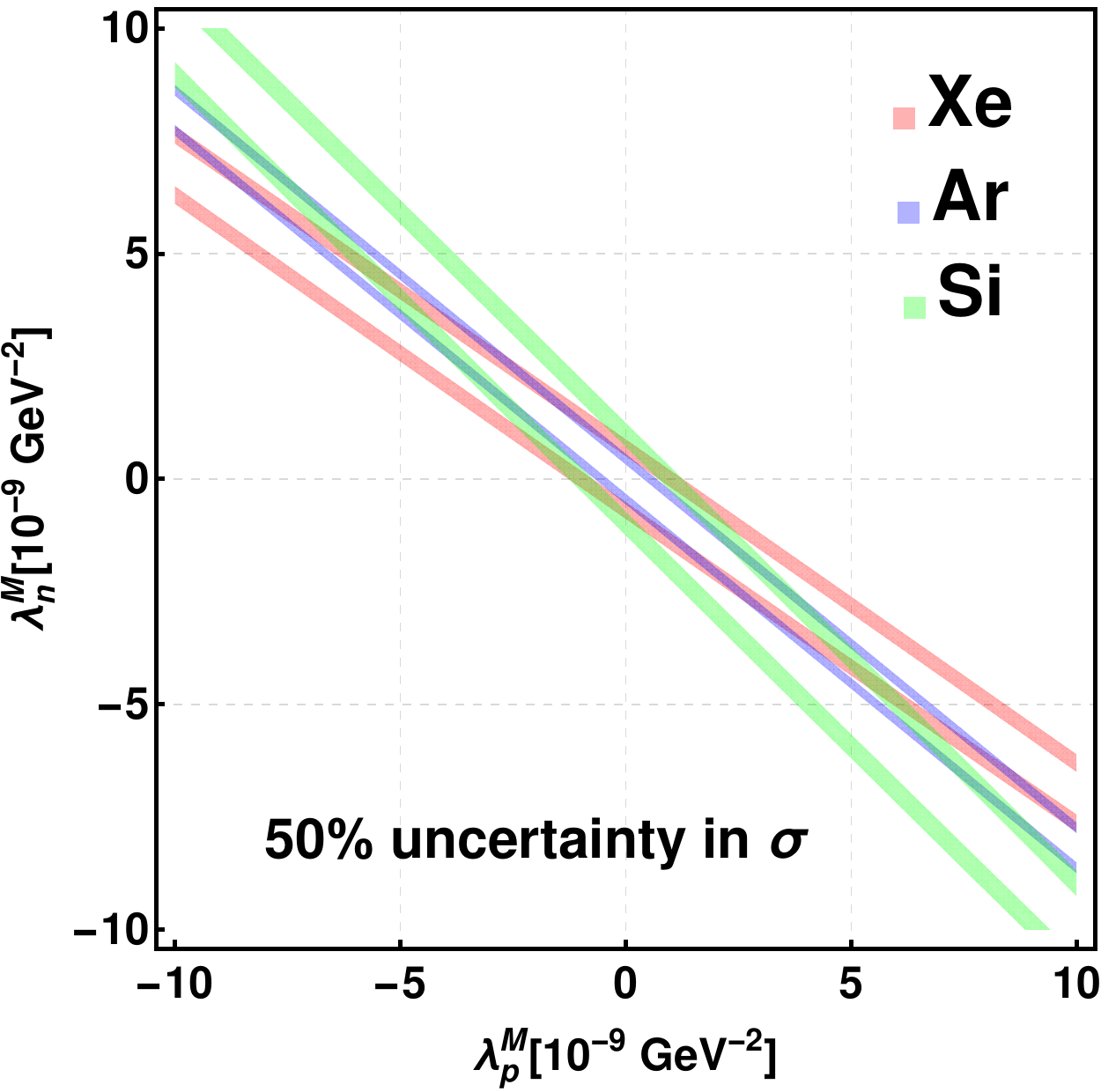}
\caption{\small The regions in the plane ($\lambda_p^M$, $\lambda_n^M$) consistent with a Majorana interpretation of the signals expected in three different targets: Xe (red), Ar (blue) and Si (green). Since these regions do not overlap, a Majorana DM particle can be excluded. The width of the bands is the result of a $50\%$ uncertainty assumed in the determination of the cross sections with the nuclei. The signal is due to a Dirac DM particle of $100$ GeV mass and  couplings $\{\lambda_{p}^D,\lambda_{p}^{\bar D},\lambda_{n}^D,\lambda_{n}^{\bar D}\}=\{6.7,2.0,-5.6,-1.0\}\times 10^{-9}\,\mathrm{GeV}^{-2}$, which, if one assumes isospin conservation (as is usually done when presenting experimental limits), corresponds to dark matter-nucleon cross sections  of \cite{Feng:2011vu} $\{0.9,0.3,1\}\times 10^{-10}\,\,\mathrm{pb}$ respectively for $\{\mathrm{Xe,Ar,Si}\}$.   \label{fig:DvsM}} 
\end{center}
\end{figure} 

Second, the targets used in DM detection experiments often include multiple isotopes of the same element. Xenon, Germanium and Silicon, for example, consist of $7$, $5$, and $3$ different isotopes with abundances larger than $1\%$, respectively, as shown in Table \ref{tab:isotopes}. For  nuclei with multiple isotopes, the event rate is thus modified to $R=\sum_i \eta_i\sigma_{X_i}I_{X_i}\approx \sum_i \eta_i\sigma_{X_i}\,I_{X}$, where the sum is over isotopes $X_i$ with fractional abundance $\eta_i$, and, in the last step,  we used the fact that the $I_{X_i}$ do not vary much for different $i$.  The cross sections discussed in the previous section should, therefore, be replaced by the weighted sum over the multiple isotopes for a given target: $\sigma_X\to \sum_i \eta_i\sigma_{X_i}$. The left-hand side of equation (\ref{eq:sigmaX}), for instance, becomes a sum of several terms, and the solution no longer consists of two parallel lines (or bands) but rather of elongated ellipses. 

Finally, we also need to ensure that the resulting cross sections are consistent with current data and that they lie, in a broad sense, within the sensitivity that could be reached by future experiments. Our task is then to find a set of three nuclei and  given values for $\{\lambda_{p}^D,\lambda_{p}^{\bar D},\lambda_{n}^D,\lambda_{n}^{\bar D}\}$ such that, under these conditions, an interpretation of the expected spin-independent signals in terms of Majorana DM can be  excluded.

Let us take $\{\mathrm{Xe,Ar,Si}\}$ as our set of three nuclei where spin-independent signals are assumed to have been observed. Xenon not only  sets the strongest limits on the spin-independent direct detection cross section \cite{Aprile:2012nq, Akerib:2016vxi} at high DM masses, but it is also the target of choice for a new generation of experiments that are currently taking data \cite{Aprile:2012zx} or are expected to run in the near future \cite{Undagoitia:2015gya}.   Argon was used by the DarkSide experiment \cite{Agnes:2014bvk} and features in some ongoing projects aiming at ton-scale detectors \cite{Aalseth:2015mba,Calvo:2015uln}.   The choice of Silicon as the third nucleus is motivated by having a ratio  $Z/(A-Z)$  close to $1$ --significantly larger than for $\mathrm{Xe}$ and $\mathrm{Ar}$-- and by the fact that it was  previously used by the CDMS collaboration as a DM target \cite{Agnese:2013rvf}.  Figure \ref{fig:DvsM} shows, for a specific parameter point,  the regions in the plane ($\lambda_p^M$, $\lambda_n^M$) that are consistent with each observation: $\mathrm{Xe}$ (red), $\mathrm{Ar}$ (blue) and $\mathrm{Si}$ (green). The width of the bands is set  by a $50\%$ uncertainty assumed in the determination of the cross sections. Even though they look like straight lines, the consistent regions for $\mathrm{Xe}$ and $\mathrm{Si}$ are actually ellipses that extend up to $|\lambda_{p,n}^M|\sim 40\times 10^{-9}\,\mathrm{GeV}^{-2}$. As expected,   the slopes of the bands are determined by the ratio $Z/(A-Z)$. We verified numerically, and it can also be directly seen in the figure, that these three regions never overlap. Consequently, the DM has to be a Dirac particle. 

Had we instead found overlapping regions, the test would have been inconclusive.  That is how the test we are proposing works. It allows, once certain conditions are met, to exclude a Majorana DM particle, being similar in spirit to the early tests suggested to test the Majorana nature of the neutrino  \cite{Rosen:1982pj,Dass:1984qc}. Summarizing, we have shown, for the first time, not only that it is possible to experimentally differentiate between Majorana and Dirac DM but also that it can be achieved with just  direct detection data.

\begin{figure}[tb!]
\begin{center}
\includegraphics[width=\columnwidth]{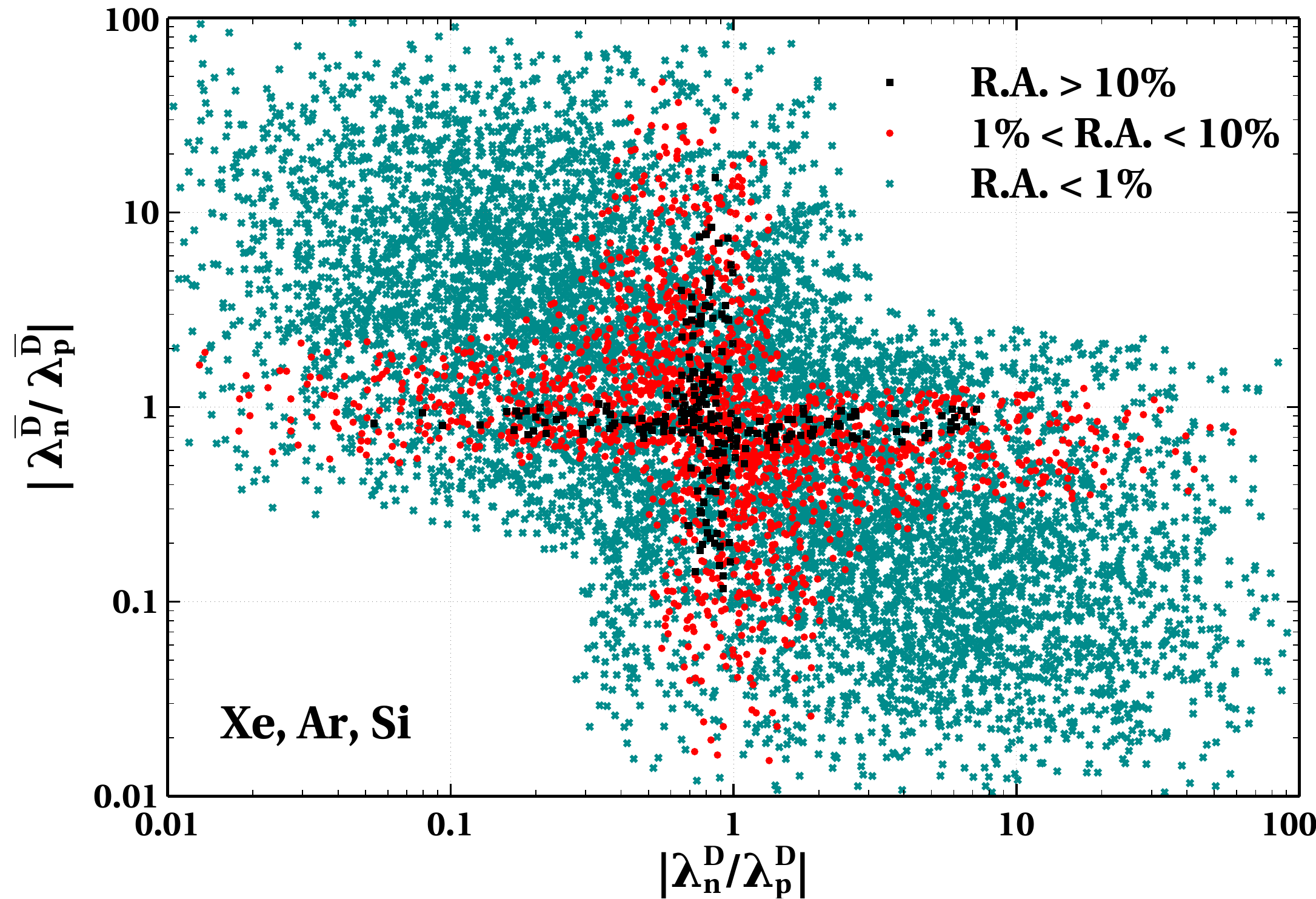}
\includegraphics[width=\columnwidth]{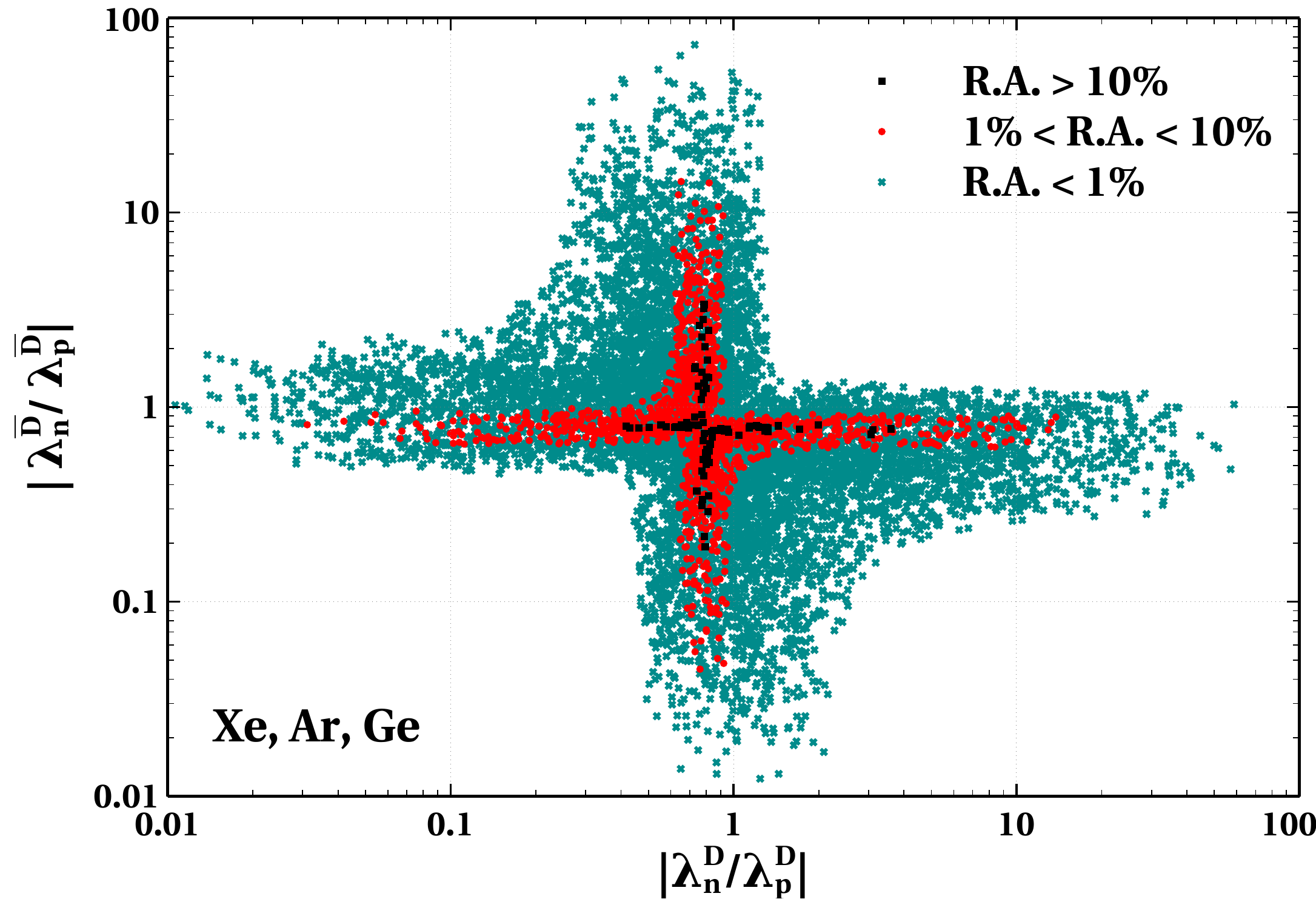}
\caption{\small Scatter plots of  $\lambda_n^D/\lambda_p^D$ versus $\lambda_n^{\bar D}/\lambda_p^{\bar D}$ for our sample of points for $\mathrm{Xe,Ar, Si}$ and $\mathrm{Xe,Ar, Ge}$ target sets. The color code indicates the required accuracy (R.A.) that enables to distinguish Dirac from Majorana DM: dark cyan for points with $\mathrm{R.A.}<1\%$, red for points with $1\%<\mathrm{R.A.}<10\%$, and black for points featuring $\mathrm{R.A.}>10\%$.   \label{fig:ScansAbs}} 
\end{center}
\end{figure} 

\begin{figure}[tb!]
\begin{center}
\includegraphics[width=\columnwidth]{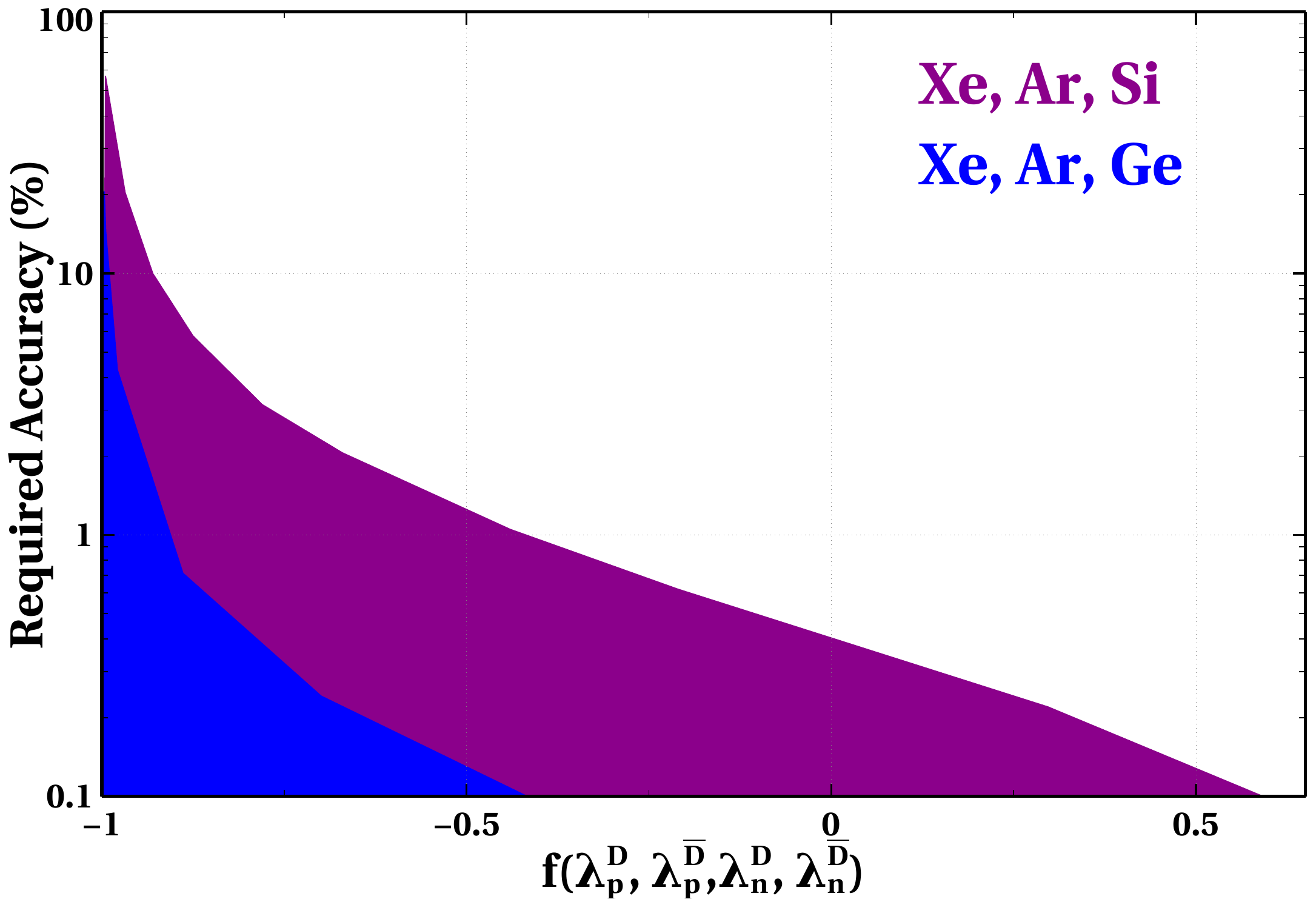}
\caption{\small The accuracy in the determination of the spin-independent cross sections required to establish the Dirac nature of the DM particle once signals are observed off $\mathrm{\{Xe,Ar,Si\}}$ (dark magenta) and $\mathrm{\{Xe,Ar,Ge\}}$ (blue).  The ordinate is the dimensionless function $f(\lambda_p^D,\lambda_p^{\bar D},\lambda_n^D,\lambda_n^{\bar D})\equiv(\lambda_p^D\lambda_n^D+\lambda_p^{\bar D}\lambda_n^{\bar D})/\sqrt{(\lambda_p^{D\,2}+\lambda_p^{\bar D\,2})(\lambda_n^{D\,2}+\lambda_n^{\bar D\,2})}$, varying from $-1$ to $+1$, which was defined to determine the accuracy needed to distinguish Majorana from Dirac Dark matter.}
\label{fig:ScansAccu}
\end{center}
\end{figure}

{\bf Detailed Analysis.} To get a better sense of how feasible it is to  distinguish Dirac from Majorana DM using the test described above, we now generalize our previous discussion in two ways. First, we consider, in addition to $\{\mathrm{Xe,Ar,Si}\}$,  a second set of three nuclei: $\{\mathrm{Xe,Ar,Ge}\}$. Germanium detectors currently provide the most stringent constraints on the scattering  cross section at low DM masses \cite{Agnese:2014aze,Agnese:2015nto}, and prominently feature in the EURECA project \cite{Angloher:2014bua}. The complementarity of $\{\mathrm{Xe,Ar,Ge}\}$, in fact, has been considered in previous works \cite{Pato:2010zk,Kavanagh:2014rya}. Regarding the discrimination between Dirac and Majorana DM, a disadvantage of $\mathrm{Ge}$  is that its ratio $Z/(A-Z)$ is not that different from that of $\mathrm{Xe}$ and $\mathrm{Ar}$. In any case, using this second set should give us an idea of how our results depend on the target nuclei. Second, we study the full parameter space relevant for spin-independent direct detection searches, which consists of $\lambda_p^D,\lambda_n^D, \lambda_p^{\bar D},\lambda_n^{\bar D}$. After randomly scanning  this parameter space
we  compute the DM cross sections off $\mathrm{\{Xe, Ar, Si(Ge)\}}$ and the accuracy required to exclude  a Majorana DM particle. The resulting sample of models is shown, projected onto different planes, in figures \ref{fig:ScansAbs} and \ref{fig:ScansAccu}, enforcing an accuracy larger than $0.1\%$.

Figure \ref{fig:ScansAbs}  displays our sample in the plane $|\lambda_n^D/\lambda_p^D|$ versus $|\lambda_n^{\bar D}/\lambda_p^{\bar D}|$. The top and bottom panel correspond respectively to the  results for the target set  $\mathrm{\{Xe, Ar, Si\}}$ and  $\mathrm{\{Xe, Ar, Ge\}}$, whereas  the colors indicate the required accuracy (R.A.) in the determination of the cross sections that enables to distinguish Dirac from Majorana DM: dark cyan for points with $\mathrm{R.A.}<1\%$, red for points with $1\%<\mathrm{R.A.}<10\%$, and black for points featuring $\mathrm{R.A.}>10\%$. Notice, in particular, that the black points --the most interesting ones-- lie just below the line $|\lambda_n^D/\lambda_p^{D}|=1$ or  $|\lambda_n^{\bar D}/\lambda_p^{\bar D}|=1$. All such points turn out to fulfill   $\lambda_n^D/\lambda_p^{D}<0$ or  $\lambda_n^{\bar D}/\lambda_p^{\bar D}<0$. That is, they all feature  a relative minus sign, and therefore a partial cancellation, between the neutron and the proton contributions to the cross section for either the DM particle or the  antiparticle.

Figure \ref{fig:ScansAccu} compares  the  accuracy required in the determination of the cross sections for the two different target sets. Since we found no points with a required accuracy larger than $60\%$ ($20\%$ for the set including $\mathrm{Ge}$), this figure suggests that future experiments should at least reach that level of precision to be able to differentiate Dirac from Majorana DM. It  also tell us that, as expected, the set  $\mathrm{\{Xe, Ar, Si\}}$ offers better prospects in this regard.  

{\bf Scalar and Vector dark matter.}
Our results so far are valid if the DM particle is a fermion, but we now show that they can be straightforwardly generalized to scalar and vector DM. If the DM particle is a scalar field ($\phi_\chi$), the effective Lagrangian describing  its spin-independent interactions with nucleons  is written as
\begin{align}
\mathscr{L}^S_{_{SI}} =& ~2\lambda_{N,e}M_\chi\, \phi_\chi^\dagger \phi_\chi\,\bar\psi_N\psi_N
\\ \nonumber
& ~+i\lambda_{N,o} \left[\phi_\chi^\dagger (\partial_\mu\phi_\chi) - (\partial_\mu\phi_\chi^\dagger) \,\phi_\chi\right]\bar\psi_N\gamma^\mu\psi_N,
\end{align}
with $\lambda_{N,o}=0$ for a real scalar. The spin-independent cross sections are given by the same expression as the Majorana (real scalar) and Dirac (complex scalar) cases --equations (\ref{eq:sigmaM}) and (\ref{eq:sigmaD}). Hence, it is possible to differentiate  real scalar DM from complex scalar DM in exactly the way already explained for fermion DM. 

For vector DM, the situation is entirely analogous --see e.g. \cite{Belanger:2008sj}.  Putting together the results for fermion DM with the generalization to scalar and vector DM just discussed, it can be concluded that the test  can potentially tell the DM particle  apart  from its antiparticle.

{\bf Outlook.} Once direct detection signals are  observed, a joint likelihood analysis would be required to   combine the data from the different experiments, to properly include all the relevant uncertainties, and to assign a precise statistical meaning to the exclusion of a self-conjugate dark matter particle.  In addition,  a more comprehensive list of possible effective operators \cite{Fitzpatrick:2012ix} could be included in the analysis.

{\bf Conclusions.} We have proposed a test that may determine if  the DM particle is different from its antiparticle. The minimum requirement to be able to do so was found to be the precise measurement of the DM spin-independent  scattering cross section  with  three different nuclei. The test consists in checking the consistency of these measurements with their interpretation in terms of Majorana  or real dark matter. If an inconsistency is found, the DM must be a Dirac or complex particle. As an illustration, we considered the target sets $\mathrm{\{Xe, Ar, Si\}}$ and $\mathrm{\{Xe, Ar, Ge\}}$, and showed that, provided the scattering cross sections are determined with a precision better than $60\%$ and $20\%$ respectively,  there exist parameter space points for which the Dirac or complex  nature of the DM particle can be established. Our results suggest that  the observation of spin-independent signals in future direct detection experiments might be used to elucidate the fundamental nature of the dark matter particle.

\section*{Acknowledgments}
We are grateful to A.\ Green and M.\ Pato for several suggestions to our manuscript. This work is supported by the DFG with grant RO 2516/6-1 in the Heisenberg program (WR), with grant RO 2516/5-1 (CY), and by the Max Planck Society in the project MANITOP (WR, CY).

\bibliography{darkmatter}

\end{document}